\documentstyle[12pt]{article}
 \advance\oddsidemargin by -1in
 \advance\evensidemargin by -1in
 \marginparsep
 0.4cm
 \advance\topmargin by
 -0.0in
 \makeindex
 \newcommand\noi{\noindent}
 \newcommand\beq{\begin{equation}}
 \newcommand\eeq{\end{equation}}
 \newcommand\beqn{\begin{eqnarray}}
 \newcommand\eeqn{\end{eqnarray}}
 \newcommand{\doublespace} {
 \renewcommand{\baselinestretch} {1.6} \large\normalsize}

\begin{document}

 \hspace*{7cm}{\large
 DOE/ER/40561-260-INT96-19-03
 \vspace*{1.5cm}
 
 \begin{center}
 
 \centerline{{\Large
 \bf
 Glauber
 Multiple Scattering Theory for}}
 \vspace{0.4cm}
 \centerline{{\Large \bf
 the Photoproduction of
 Vector Mesons off Nuclei}}
\vspace{0.5cm}
 \centerline{{\Large \bf and the Role of the
 Coherence
 Length}}
 
 \vspace{0.5cm}
 
 \bigskip
 
 {\large J\"org~H\"ufner}\\
\medskip
 {\sl Institut f\"ur Theoretische Physik der Universit\"at\\
 Philosophenweg 19, D-69120 Heidelberg, Germany}
 \bigskip
 
 {\large
 Boris~Kopeliovich}\footnote{On leave of absence from
 Joint Institute for
 Nuclear
 Research, Laboratory
 of Nuclear Problems,\\ Dubna, 141980 Moscow
 Region,
 Russia. E-mail:
 bzk@dxnhd1.mpi-hd.mpg.de}
 
 \medskip
 
 {\sl Institute for Nuclear Theory, University of Washington\\
 Seattle, WA98195, USA}\\
 and\\ 
 {\sl
 Max-Planck-Institut f\"ur Kernphysik\\ 
 Postfach 103980,
 D-69029 Heidelberg,
 Germany}
\bigskip

{\large Jan~Nemchik}\footnote{On leave of absence 
from Institute of Experimental Physics
 SAV, Solovjevova 47, CS-04353 Kosice, Slovakia}\\
\medskip
{\sl Dipartimento di Fisica Teorica, Universit\`a di Torino\\ I-10125, Torino,
 Italy} 
 
\end{center}

\vspace{.5cm}
\begin{abstract}

 The integrated cross section for the incoherent  photoproduction
 of vector mesons on nuclei $\gamma^* A \to VX$,
 $X\not=A$,
 is
calculated within Glauber theory and as a function of the photon
 energy. The
inverse of the longitudinal momentum
 transfer is called coherence length $l_c$ and depends on the 
virtuality and the
energy
 of the photon. Nuclear
transmission factors strongly depend on
 $l_c/R_A$ ($R_A$ is the nuclear
radius) and this effect may interfere
 with the search for color transparency
effects.
 
\end{abstract}

\newpage
 \doublespace
 
 With the help of the new powerful accelerators
 like CEBAF and HERA,
 which produce high energy (real or virtual) photons,
 the investigation
 of the hadronic processes inside nuclei (like color
 transparency, for
 example) will make substantial progress.  Photons are
 particularly
 suitable for these studies since the production vertex for
 hadronic
 probes inside nuclei is well known and the analysis
 of
 the results is simpler and more reliable.  In this note we treat the
 photoproduction of vector mesons inside nuclei and their 
rescattering
 via
 strong interactions within Glauber multiple scattering theory 
\cite{glauber}.
 This is a
 simple problem treated by a conventional well accepted
 method.  Yet,
 interference effects, which have not been treated before
 lead to a
 non-trivial energy- and $Q^2$-dependence of the production
 cross section and
 to unexpected phenomena.  Any search for
 unconventional effects, like color
 transparency, should should refer
 to our calculations as a base line.
 
 We study the incoherent photoproduction of
 a
 vector meson $V$ on a nucleus
 $A$, $\gamma^*A \to
 VX$, where X is any
 state of the target nucleus except the
 ground state
 $|0\rangle$. The transition
 operator is given in eikonal form
 as
 \beq
 \Gamma^{\gamma V}_A(\vec b;\{\vec s_j,z_j\}) =
 \sum_{j=1}^A
 \Gamma^{\gamma V}_N(\vec b-\vec s_j)\
 e^{iq_Lz_j}\
 \prod_{k(\not=j)}^A\left[1-
 \Gamma^{VV}_N(\vec b-\vec s_k)\
 \Theta(z_k-z_j)\right]\ ,
 \label{1}
 \eeq
 where $\{\vec s_j,z_j\}$
 denote
 the coordinates of
 the target nucleon $N_j$. $\Gamma^{\gamma V}_N$
 is the
 vector meson photoproduction amplitude on a nucleon,
 while
 $\Gamma^{VV}_N$ describes the
 elastic
 scattering amplitude for the vector
 meson on a
 nucleon. The
 $\Theta$-function allows vector
 meson
 rescattering only on those nucleons
 $k$,
 which lie ``behind'' the
 production point, i.e. $z_k >
 z_j$. The momentum
 $q_L$, which appears in
 the
 phase factor in eq.~(\ref{1}) is the difference
 between the
 longitudinal momenta of the photon and
 of the produced vector meson
 \beq
 q_L=p^{\gamma} - p_V^L =
 \frac{Q^2+M_V^2+(p_V^T)^2}
 {2\nu}\ .
 \label{2}
 \eeq
 The quantity $q_L$, which depends on the energy
 $\nu$ and
 the virtuality $Q^2$ of the photon plays the central role in our 
consideration.
 Its
 inverse $l_c=1/q_L$ is called the coherence length.  Because of the
 phase factor in eq.~(\ref{1}), the photon production amplitude on 
two
 nucleons with positions $|z_i-z_f| < l_c$ add up coherently, otherwise
 there is
 destructive interference.
 
 The incoherent cross section for the production
 of a vector meson with
 transverse
 momentum $p_V^T$ in a reaction, where
 the nuclear state changes from
 $|0\rangle$ to $|f\rangle$ ($f\not=0$) is
 given in Glauber theory by
 
 \beq
 \frac{d\sigma^{\gamma V}_{inc}(0\to f)}
 {d^2p_V^T} =
 \left |
 \int\frac{d^2b}{2\pi}\ \exp(-i\ \vec p_V^T\ \vec b)\
 \left\langle f
 \left|\Gamma^{\gamma V}_A(\vec b) \right |0\right\rangle
 \right |^2\ ,
\label{3}
\eeq
 
 \noi
 from which one obtains the total incoherent cross
 section
$\sigma^{\gamma V}_{inc}$ by integrating over $\vec p_V^T$ and
 summing over
all final states $|f\rangle$ with the exclusion of the
 ground state:
 
 \beq
 \sigma^{\gamma V}_{inc} = \int d^2b\
 \left[\langle 0|
 \left|\Gamma^{\gamma V}_A(b)\right |^2 |0\rangle -
 \left |\langle 0|\Gamma^{\gamma
V}_A|0\rangle\right|^2\right]\ .
\label{4}
\eeq
 
 \noi
 We introduce the simplifying assumption
 
 \beq
\Gamma^{\gamma V}_N(b) = \lambda_V\ \Gamma^{VV}_N(b)\ ,
\label{5}
\eeq
 
 \noi
 where $\lambda_V$ is a constant, which equals to $e/f_V$ in the
vector
 dominance model. Eq.~(\ref{5}) is a good approximation as long as
the widths of the profile functions $\Gamma^{\gamma V}_N$ and
 $\Gamma^{VV}_N$
are small compared to the nuclear radius.  We
 also use the relations for the
elementary scattering processes of a
 vector meson on a nucleon
 
 \beqn
 &
&{1\over 2}\sigma^{VN}_{tot} = \int
 d^2b\ Re\ \Gamma^{VV}_N(b)\ ,\\
\nonumber
 & &\sigma^{VN}_{el} = \int
 d^2b\ \left|
\Gamma^{VV}_N(b)\right|^2\ .
\label{6}
\eeqn
 
 \noi
 Furthermore, the nuclear wave function $|0\rangle$
 is
assumed to be a product of single particle wave
 functions and to be
completely described by the
 density distribution $\rho(\vec r)$, which is
normalized to $A$.  In order to bring out the
 essential effects, which are
associated with the
 coherence length $l_c=1/q_L$, we evaluate the
incoherent cross section eq.~(\ref{4}) for two
 limiting cases, $q_L \to 0$
(high energy) and
 $q_L \to \infty$ (low energy), before we give the
 exact
expression.  For both limiting cases simple
 analytical formulae can be
derived.
 
 In the high energy limit, $q_L \to 0$, the phase
 factor
$\exp(iq_Lz)$ can be dropped in
 eq.~(\ref{1}), and after some algebra one
finds
 
 \beq
 \Gamma^{\gamma V}_A(\vec b;\{\vec s_i\})
 =
\lambda_V\left\{1-
 \prod_j\left[1-\Gamma^{VV}_N(\vec b-\vec
s_j)\right]\right\} = \ \Gamma^{VV}_A(\vec
 b;\{\vec s_i\})\ ,
\label{7}
 \eeq
 
 \noi
 where $\Gamma^{VV}_A$ is the Glauber amplitude for
 the
scattering of the vector meson $V$ by
 the nucleus $A$.
 Using
eqs.~(\ref{4}),~(6) and (\ref{7}), the incoherent cross
 section can be 
evaluated to give
 
 \beq
 \sigma^{\gamma V}_{inc} = \lambda_V^2\ \int d^2b\
\left[e^{-\sigma^{VN}_{in}T(b)} -
 e^{-\sigma^{VN}_{tot}T(b)}\right]\ ,
\label{8}
 \eeq
 
 \noi
 where $T(b)=\int_{-\infty}^{\infty}dz\ \rho(b,z)$
 is the
nuclear thickness function, and $\sigma_{in}^{VN}
 = \sigma_{tot}^{VN} -
\sigma_{el}^{VN}$.
Note that eq.~(\ref{8}) has, up to the factor $\lambda_V^2$, the same  
form as the expression (see in \cite{tarasov})  
for the cross section of quasielastic
scattering of a hadron ($V$) on a nucleus.
 In the low energy limit, when $q_L \gg R_A^{-1}$
 ($R_A$ being the nuclear radius) all expressions which explicitly 
contain
 the phase factor go
 to
zero, e.g.
 
 \beq
 \left\langle 0\left|\Gamma^{\gamma
V}_A\right|0\right\rangle = 0\ .
 \label{9}
 \eeq
 
 In the expression
\beqn
 & &
 \langle 0|\left|\Gamma^{\gamma
 V}_A\right|^2|0\rangle =
\lambda_V^2 \left\langle 0\left|\sum_{i,j}
 \Gamma^{VV}_N(\vec b-\vec
 s_i)\
\Gamma^{VV^*}_N(\vec b-\vec s_j)\
 e^{iq_L(z_i-z_j)}\ \times
 \right.
\right.
 \nonumber
 \\
 & &
 \left.
 \left.
 \prod_{k(\not
 =i)}\left[1-
\Gamma^{VV}_N(\vec b-\vec s_k)
 \Theta(z_k-z_i)\right]
 \prod_{l(\not
=j)}\left[1-
 \Gamma^{VV^*}_N(\vec b-\vec s_l)
 \Theta(z_l-z_j)\right]
\right|0\right\rangle
\label{10}
 \eeqn
 
 \noi
 all terms with $i\not = j$ do not contribute
 to the
matrix element and one is left with
 
 \beq
 \sigma^{\gamma V}_{inc} =
\lambda_V^2\ \sigma^{VN}_{el}\int d^2b\
 \int\limits_{-\infty}^{\infty}dz\ \rho(b,z)\
e^{-\sigma^{VN}_{in}T_{z}(b)}\ ,
 \label{11}
 \eeq
 
 \noi
 where $T_z(b)
= \int_{z}^{\infty}dz'\rho(b,z')$.
 
 The low and high energy 
limits,
eqs.~(\ref{11}) and (\ref{8}), respectively,
were derived also in \cite{kz}.  
In order to expose the difference 
between these expressions
 we consider the
case where the elastic $VN$ cross
 section is small $\sigma^{VN}_{el}T(b)\ll
1$.
 Then we find
 
 \beq
 $$
 \begin{displaymath}
 \sigma^{\gamma V}_{inc} =
 \sigma(\gamma^*N\to
 VN)\ \int d^2b\ 
 \int\limits_{-\infty}^{\infty}
 dz\ \rho(b,z)\ \times
 \left\{
\begin{array}{ccc}
 \ e^{-\sigma^{VN}_{in}T_{z}(b)} & ({low\
 energy})&\\ &
 &\\
 \
 e^{-\sigma^{VN}_{in}T(b)} & ({high\ energy})\
 ,&
\end{array}
 \right.
\end{displaymath}
 $$
 \label{12}
 \eeq
 
 \noi
 where $\sigma(\gamma^*N \to VN) =
 \lambda_V^2\
 \sigma^{VN}_{el}$ is the photoproduction cross
 section on a
 nucleon.  The two limiting cases, for
 low and high energies, respectively,
 differ in the
 attenuation factors.  The integration in eq.~(12) goes over
 all points $(b,z)$ for the photoproduction vertex $\gamma^*N \to 
VN$.
 In
 the low energy limit, the attenuation of the 
outgoing vector meson is governed
 by the
 $T_z(b)$, which is the nuclear thickness
 experienced by the vector meson
 from the point of
 creation $(\vec b,z)$ until its exit from the nucleus.
 This
 is what is expected.  

At high energy, on the
 other hand, the attenuation is
 governed by $T(b) =
 T_{z=-\infty}(b)$, which is the thickness of the
 nucleus along the total path (the summed paths of
 the photon and the vector
 meson).  The attenuation
 is governed by the same $\sigma_{in}^{VN}$
 independently of whether it acts on the incoming photon or the
 outgoing vector meson.  We have to
 conclude that at high energies the photon has
 already converted to a virtual vector meson long
 before the vector meson is
 put on the energy shell
 via an interaction with a target nucleon at
 $(\vec b,z)$. This interpretation is familiar from the
 vector dominance model (VDM)
 \cite{bauer}, which
 predicts that at high energies a photon behaves in
 strong interactions like a hadron.  Or the same phenomenon 
expressed
 in quantum mechanical language: because of the uncertainty 
principle a photon can
convert
 virtually into a vector meson. The lifetime $\Delta t$ of
 such a fluctuation in the lab. frame 
 is estimated to be $\Delta t \sim
 1/q_L$. If $\Delta
 t \gg 1/R_A$ the chances are large that the
 incoming
 photon is already in the virtual state of
 a vector meson and therefore
 experiences a strong
 absorption {\it before} the fluctuation is put on the
 energy shell.
 
 Note that we
 have not used the vector dominance model, but only eq.~(\ref{5}),
 which
 is rather general.  However, the laws of quantum mechanics have
 been properly accounted for in eq.~(\ref{1}).  Taking all the 
interference
 terms in the multiple scattering
 series into
 account leads to the different attenuations at low
 and high
 energies, i.e.  for $q_L \to \infty$ and
 $q_L \to 0$, respectively.
 
 What is the proper scale, which separates high and
 low energies? There are a two
 dimensioned scales in the
 multiple scattering series, the nuclear radius
 $R_A$, and the mean free path of the vector meson in nuclear matter 
 $l_{free}\approx 1/\sigma_{in}^{VN}\rho_0$ ($\approx 3\ fm$ for
 $\rho-meson$). 
 Condition $l_c \ll L_A=\min\{R_A;l_{free}\}$ defines low
 and $l_c \gg L_A$ high
 energies, where  $l_c = 1/q_L$ can be interpreted as the
length, which a quantum mechanical
 fluctuation
 $\gamma \to V \to \gamma$ travels if
 it has the speed of light. We call $l_c$
 the {\it coherence length}.
 
 The
general expression for the incoherent
 photoproduction cross section on a
nucleus can be
 also calculated from the  expression
 eq.~(\ref{4}).
The result is a somewhat lengthy
expression
 
 \beqn
 \sigma^{\gamma V}_{inc} &=&
 \sigma(\gamma N\to VN)\
\int d^2b\left\{
 \int\limits_{-\infty}^{\infty}
 dz\ \rho(b,z)\
e^{-\sigma^{VN}_{in}T_z(b)}\ +
 \right.
 \nonumber\\
 & &
 {1\over 2}\
\frac{\sigma^{VN}_{tot}}
 {\sigma^{VN}_{el}}\
 (\sigma^{VN}_{in} -
\sigma^{VN}_{el})\
 \int\limits_{-\infty}^{\infty} dz_1\
 \rho(b,z_1)\
\int\limits_{z_1}^{\infty} dz_2\
 \rho(b,z_2)\
 \times \nonumber\\
 & &
\cos\left[q_L(z_2-z_1)\right]\
 \exp\left[- {1\over 2}(\sigma^{VN}_{in} -
\sigma^{VN}_{el})\ T_{z_2}(b) -
 {1\over 2}\sigma^{VN}_{tot} T_{z_1}(b)
\right]\ -
 \nonumber \\
 & &
 \left.
 {1\over 4}\
\frac{\left(\sigma^{VN}_{tot}\right)^2}
 {\sigma^{VN}_{el}}\
 \left|
\int\limits_{-\infty}^{\infty}
 dz\ \rho(b,z)\ e^{iq_Lz}\
 e^{-{1\over
2}\sigma^{VN}_{tot}\ T_z(b)}
 \right|^2 \right\}\ ,
\label{14}
\eeqn
 
 \noi
 where the last term in the curly brakets is the cross section
for
 coherent production $\gamma A \to V A$ \cite{bauer}.
 Formula (\ref{14}) for incoherent photoproduction
 is new and is the main result of the present paper.
 
 In the two limits $q_L \to
0$ and $q_L \to \infty$ formula(\ref{14})
 leads to expressions (\ref{8})
and (\ref{11}),
 respectively, as it should be.
 As an illustration we
show numerical examples for the effect of the
 coherence length. We calculate
the nuclear transmission function
 $Tr^{\gamma V}(q_L,A)$, which is defined
by
 
 \beq
 Tr^{\gamma V}(q_L,A) =
 \frac{\sigma_{inc}(\gamma A \to
VX;q_L)}
 {A\ \sigma(\gamma N \to VN)}\ ,
\label{15}
\eeq
 
 \noi
for the incoherent photoproduction of 
$\rho$-mesons as a function of the energy. The results for carbon, 
iron and lead are
shown  in Fig.~1 (one can find more examples in \cite{kn}) 
as a function of the energy $\nu$ and the virtuality 
$Q^2$ of the
photon. Both quantities appear in $q_L$ eq.~(\ref{2}).

The transverse momentum $p^T_V$ in eq.~(\ref{2}) has been set 
equal zero. The
cross sections which enter the calculation are 
$\sigma^{\rho N}_{tot}=25\ mb$ and
$\sigma^{VN}_{el}= (\sigma^{VN}_{tot})^2/16\pi B^{VN}_{el}$, where
$B^{\rho N}_{el}\approx 8\ GeV^{-2}$ is the 
slope of the differential cross section
of elastic $\rho N$ scattering. 

\begin{figure}[tbh]
\includegraphics{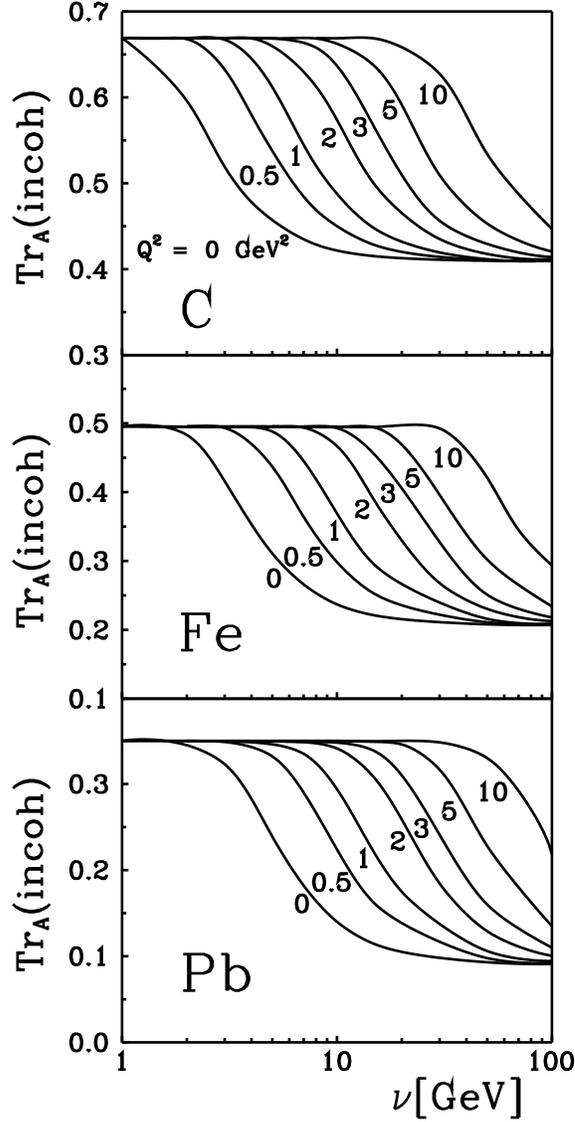}
\begin{center}
\vspace{14.5cm}
\parbox{13cm}
{\caption[Delta]
{Nuclear transparency for incoherent photoproduction
of $\rho$-mesons on $C,Fe$ and $Pb$ nuclei as a function of the 
photon energy $\nu$
and for values $Q^2$ of the virtuality of the photon between 0 and 
10~(GeV/c)$^2$.}
\label{fig1}}
\end{center}
\end{figure}

The numerical results in Fig.~1 display a strong variation of the 
nuclear
transparency as a function of $\nu$ and $Q^2$. The numerical 
values in the low and
high energy limits ($q_L\to\infty$ and $q_L\to0$, respectively 
obey the approximate
relation
 \beq
 Tr(q_L \to 0)
\approx [Tr(q_L \to \infty)]^2.
 \label{16}
 \eeq
 
 The half value between the low and high energy limits is reached 
for $(q_L\cdot
R_A)\simeq1.5$, where $R_A=1.2 A^{1/3}$~fm. 
 
 Although eq.~(\ref{14}) can be evaluated numerically in a rather 
straightforward
way, as is done for Fig.~1, 
we have also looked for approximate expressions for the 
nuclear transparency.
In the limit of small attenuation, $\sigma^{VN}_{in}\cdot\langle 
T\rangle \ll1$, the
 transparency
becomes
 \beq
 Tr^{\gamma V}(q_L,A)\approx\left\{1 - {1\over 
2}\sigma^{VN}_{in}\
 \langle
T\rangle\
 \left[1+F^2_A(q_L)\right]\right\}\ ,
 \label{17}
 \eeq
 where
$F_A$ is the nuclear formfactor
 \beq
 F^2_A(q_L) = {1\over A\ \langle T\rangle}\ \int
 d^2b\ \left|\int\limits_{-\infty}^{\infty}dz\ 
\rho(b,z)\
 e^{iq_Lz}\right|^2\ ,
 \label{18}
 \eeq
 and
 \beq
 \langle
T\rangle =
 {1\over A}\ \int d^3r\ \rho(\vec r)\
 T(b)
\label{19}
 \eeq
 is the mean nuclear thickness. For $q_LR_A \gg 1$
 (low
 energy) one
 has $F_A \to 0$, but $F_A \to 1$
 if $q_LR_A \ll 1$ (high
 energy). The
 corresponding
 transparencies differ by a factor of two in the
 second term
 in the square brackets in
 eq.~(\ref{17}). The approximate formulae may be good for $J/\psi$ 
photoproduction as it was suggested in \cite{benhar},
but fails for the $\rho$-meson.
 
 We have presented a fully quantum mechanical derivation for the 
incoherent
photonproduction of vector mesons off nuclei using Glauber's 
formalism of multiple
scattering theory. Strong energy and $Q^2$ dependencies are 
predicted, which arise
from interference effects, which essentially depend on the 
coherence length $l_c=1/q_L$.
 For low energies, $l_c \to 0$, one 
recovers the
classical picture of a vector meson produced inside the nucleus and 
attenuated on
its way out. In the other limit of high energies, $l_c \to \infty$, 
the incoming
photon is already a virtual vector meson and thus experiences 
attenuation before it
converts to the on-shell meson. This strong dependence on $q_L$
leads to a rising $Q^2$-dependence at fixed energy (see examples in
\cite{kn}) and has to be accounted
for before one may look for effects like color transparency.

 {\bf
 Acknowledgement:} The work was partially
 supported by a grant from the Federal Ministry for Education and 
Research (BMBF), grant number 06~HD~742, and by INTAS grant 93-0239.
B.K. thanks the Institute for Nuclear Theory 
at the University of Washington
for its hospitality and the Department 
of Energy for partial support during the
completion of this work.

 \end{document}